# A Detail Study of Security and Privacy issues of Internet of Things

Mohan Krishna Kagita[*], Navod Thilakarathne, Dharmendra Singh Rajput, and Dr. Surekha Lanka

[1] School of Computing and Mathematics,Charles Sturt University, Melbourne, Australia.
[2] Department of ICT, University of Colombo,Sri Lanka.
[3] School of Information Technology and Engineering, Vellore Institute of Technology, India.
[4] IT Department, Director of Business & Technology, Stamford International University, Bangkok, Thailand.
mohankrishna4k@gmail.com,navod.neranjan@ict.cmb.ac.lk,
dharmendrasingh@vit.ac.in,surekha.lanka@stamford.edu

**Abstract.** The Internet of Things, or IoT, refers to the billions of physical objects around the planet that are now connected to the Internet, many of which store and exchange the data without human interaction. In recent years the Internet of Things (IoT) has incredibly become a groundbreaking technical innovation that has contributed to massive impact in the ways where all the information is handled in corporate companies, computer devices, and even kitchen equipment and appliances, are designed and made. The main focus of this chapter is to systematically review the security and privacy of the Internet of Things in the present world. Most internet users are genuine, yet others are cybercriminals with individual expectations of misusing information. With such possibilities, users should know the potential security and privacy issues of IoT devices. IoT innovations are applied on numerous levels in a system that we use daily in our day-to-day life. Data confidentiality is a significant issue. The interconnection of various networks makes it impossible for users to assert extensive control of their data. Finally, in this chapter discusses the IoT Security concerns in the literature and providing critical review on current approach and proposed solutions on present issues on Privacy protection of IoT devices.

## 1 Introduction

The present status of the world is loaded up with immense innovation types of technologies [1]. Innovative manifestations show up each day.

---
[*]

The introduction of the Internet of Things accompanied supreme improvements where they filled in different businesses. As of now, universal frame- works are equipped for various types of cooperation[36]. They are made out of interconnected systems to share and get data. Undoubtfully, IoT has genuinely enhanced individuals' lives. Worldwide measurements demonstrate the utilization of IoT will ascend later on [37]. New real factors are target giving apparent helpfulness of the IoT gadgets. Because of the broadened interconnectivity, the principal challenge lies in the middle of security and privacy[11]. IoT is combined with several layers of connectivity. Each area is shouting out for top to bottom security and protection measures forestalling unapproved misuses. Regardless of the apparent capacities of IoT, there is a generous hole with regards to the accomplishment of sufficient security and privacy measures. The present difficulties being confronted will end up being increasingly extreme later on. Aggressors consistently ad-lib test procedures to demonstrate their theory [14]. There are vulnerabilities with respect to the idea of IoT security and protection issues. In this way, much consideration ought to be committed to genuineness, secrecy, and information uprightness in IoT devices.

## 2   An Overview of the IoT Market

There is a worldwide rise in the status of IoT devices. Anticipation is made for the global reach of IoT devices. Statistical reviews indicate more than 1.1 trillion dollars will be the market value for IoT devices by 2026[12]. The report by Fortune Business Insights also highlighted the global market of IoT in 2018 'valued at $190 billion'. Out of these, the banking and financial sector carries the day with high-value IoT inputs. Globally, banking, financing, and insurance services capped at USD 17.85 billion years ending 2018. The healthcare industry also faces great composures among the developments of the IoT[17]. Zion Market Research's data valued IoT healthcare at $14 billion by 2024. They attributed the rise to the widening growth of the cloud in healthcare operations.

Further, there is a growing concern about the number of IoT devices available worldwide. It is indicative the widening utilization of IoT gives the rise of these devices in the future. On the same issue, predictive statistics are on the course. Gartner claimed about 14.2 billion IoT devices are actively in use at the close of 2019. Anticipations on the same predicted over 25 billion devices would be in place by 2025[13]. It is a figure supporting their prior estimations. With the current trends, it is uncertain to predict the future growth of IoT accurately. Evidence suggests the growth of the industry enormously surpassing the already known.

IoT adoption and success rates are some of the significant figures supporting endless growth. IoT deployment in financial institutions exceeds a 58 percent success rate [1]. A report by Forbes Insight Survey reported well-developed IoT initiatives. Deloitte surveyed Industry 4.0's executives from 11 nations and said 94 percent of their digital transformation among their strategic initiatives[35]. Gartner studies from various organizations reported an 80 percent success rate. Those organizations reported better results exceeding their expectations. Further, a report from International Data Corporation (IDC) from 29 nations stated 85 percent of budgets for the IoT basis. Relative insights from IoT applications are heavily dependent on industrial revolution 4.0. The status of the matter indicates that almost everything is connected to the internet[15]. Studies focus on the forward momentum. IoT applications will rise in the future, where companies continue to strive for virtual connections[5]. These transformations are regardless of the number of people reading on the same page.

## 3 Data Analysis of IoT Security and Privacy

### 3.1 Size of IoT Market

The size of the market remains essential as a tool for adequate plans and preparations[34]. Over years, there is a growing number of IoT devices worldwide. These devices work in various sectors. Each of the specific industry bears multiple forms of IoT devices. Similarities of these device exist. Companies strive on making superior devices surpassing their competitors to increase perceived usefulness.

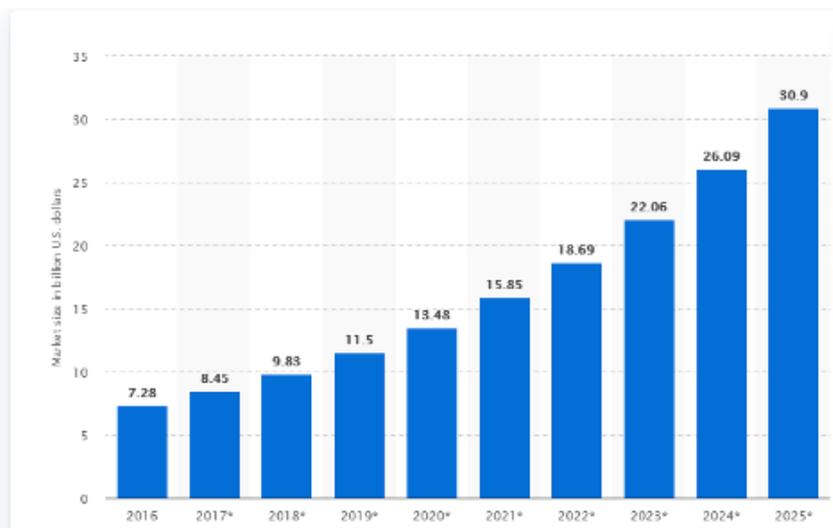

**Fig. 1.** Chart. Estimated Size of IoT Market. Source Statistica.com

The analysis presents the current status of the IoT market. Based on data gotten from 2016, the market is exponentially growing[38]. Each year serves as an increment of the current devices available in the market. In the previous year, the market size was estimated to be 11.5 billion dollars. Based on predictive statis- tics, the figure will rise in the future. New creations in the market attribute to enhanced investments in the sector[33]. Businesses, governments, and other investors continue to inject more resources into the market.

### 3.2   Spending on IoT

As earlier mentioned, the initial call for rapid developments in IoT is through improvised ways of spending. Major investors and stakeholders in the industry continue to own significant events for the case[32].

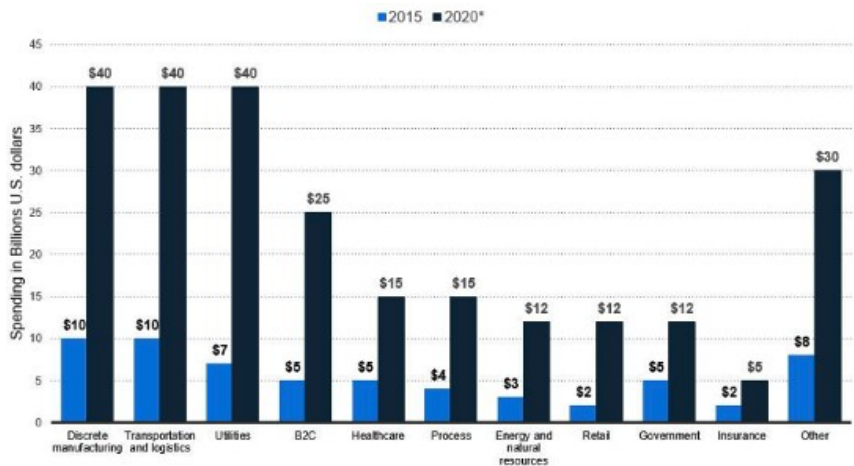

**Fig. 2.** Spending on the IoT by Industry (2015 vs 2020). Source Forbes

According to a statistical review by Forbes, the global spending on IoT is overwhelming [30]. The chart represents various industries supposedly utilizing IoT devices. Based on the reviewed map, the assumptions vary based on the comparison of the two years. Manufacturing, transportation, and utility industries are on a rapid shift in terms of IoT utilization. Business to business applications are also real Other upcoming players include the healthcare industry, which is experiencing rapid changes now and then. The trend is indicative of the responses of multiple industries concerning IoT[31].

## 4   Classification of IoT Security Issues

A rating by Threatpost.com indicated various forms of attacks. Three major categories of security and privacy issues underlie[39]. The first thing of concern is the exploits. They lead by 41 percent of the total problems. Exploits encompass several activities by cybercriminals[21]. These are activities conducted by criminals without users attempt. Holding such exploitations means to gain access to the IoT systems. Attackers will attempt to alter the normal operations of the system. Various classifications are under the exploit niche. Some of these include command injections, remote code execution, and SQL injections. The other major classified issue is malware[45]. Malware is very capable when it comes to the threat of IoT security and privacy. According to the study, malware accounts for 33 percent of these threats. Worms are the most substantial portion of malware. Other significant issues include ransomware, backdoor trojan, and botnet in the order. User practices are identified to be a significant issue at stake. These are the general practices of individuals. People are often peculiar during device usage. As mentioned, some users bear particular reasons for using the devices. User behavior could bring significant security and privacy issues. Passwords are some of the major causes of security and privacy issues[40]. Weak passwords are insecure and could be easily cracked. Most IoT devices prompt the use of passwords to access these services. Other problems with user practice include crypto jacking and phishing[41].

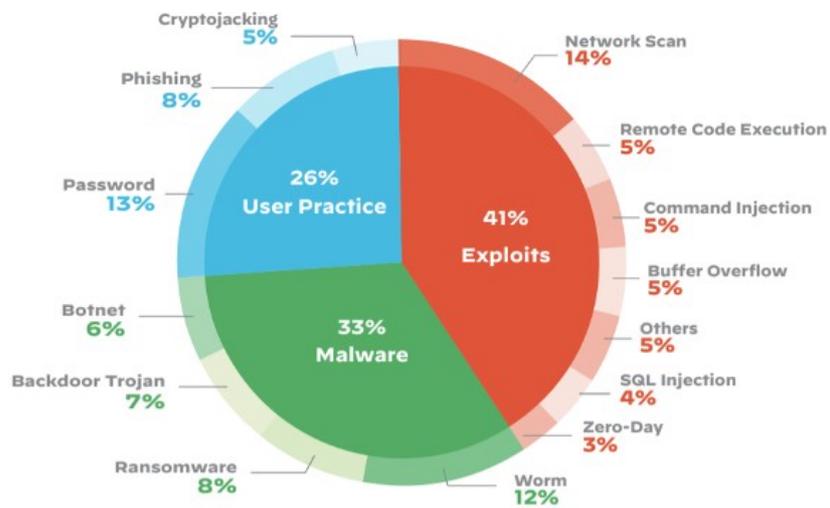

**Fig. 3.** Chart. IoT Security and Privacy Issues by Severity. Source Threatpost.com

## 5 IoT Security and Privacy Issues

IoT devices include interconnectivity starting with one point then onto the next. The vast majority of them are not self-sufficient elements[44]. They include connections using across the internet, encounters several loopholes weakening security and privacy apparatus. Security specialists set out on different exercises to authenticate the issues encompassing IoT gad- gets. Digital assailants newly direct assaults in the wake of confirming powerless focuses inside these devices. The way is relevant with regards to security examination. As of late, ethical hacking is broadly used as an idea of getting security and protection escape clauses [42]. It is where uncommon security specialists take up the job of 'genuine hackers' and perform tests to bear wit- ness to for issues. Ethical hacking gets uplifted examinations from PC security subtleties. Such notable acts bring a nearby acknowledgment of the problems of IoT security and protection. At the point when escape clauses are distinguished, better countermeasures are set up to contain them.

Enormous stress is the degree of IoT security and privacy. Besides, it shows up across different devices extraordinarily. Similarities still exist in a number of these security and privacy issues [6]. But mitigation plans always vary based on several parameters like the programing language and the use of the device. Distinct examinations continue to prove security countermeasures for every problem.

Increasingly confused security and privacy issues will bring up later on. In any case, the crucial items are giving security specialists restless evenings endeavoring to make a suffering arrangement. In this way, there is a need to diagram a portion of these shared traits influencing IoT security and privacy.

### 5.1 Lack of Compliance

Compliance is one of the significant factors supposed to solve underlying IoT se- curity and privacy issues. The widening growth of IoT applications is widening every day[19]. New devices are born now and then. Many players in the industry are anonymous. While strict regulations are still in place, there is a widening rift in attempting to take control of all the IoT developments. Countries across the globe establish governing bodies to create appropriate frameworks for computer applications. For example, the Institute of Electrical and Electronics Engineers (IEEE) withholds their advanced theories, practice, and the overall appeal of technologies[7]. Compliance within the computer society purposes the right inputs to be included in the making of these developments [27]. Undiscovered developers anonymously sharing their IoT creations bear diminished inputs in security and privacy apparatuses. Failure to acknowledge international security standards raises the chances of security loopholes. The end products are problematic faced with several issues like inappropriate hardware, lack of security updates, and insecure transfers.

### 5.2 User Awareness and Knowledge

The growing use of the internet paves the way for awareness and knowledge [2]. Over the years, there is a growing number of factors leading to inappropriate security measures. Awareness and knowledge account for responsible individuals when using technological aspects. IoT is a new technology with complicated bits. Priorly, people were fond of acknowledging security measures over the internet. It is common to distinguish between a real email and spam. Virus scans are also frequent among internet users. When it comes to the IoT applications, there are complications[26]. Most people are unaware of the real demands of these devices. Security and privacy issues primarily lie on the manufacturers' side. However, perceived users constitute a significant call for alarm. Lack of awareness about the workability of the IoT devices leaves the users channeled to security and privacy risks. Ignorance also works under the context when it comes to the functions of the devices. Attackers often use social engineering, tricking users [8]. Here, they find meaningful information directing them to critical resources.

### 5.3 Lack of Testing and Device Updates

Before the release of the IoT devices, regular tests should be conducted to ascertain their effectiveness. The anonymity of these devices creates ignorance for the creators. They are unable to test actively and update their devices to the latest security patches. Such levels of carelessness leave IoT devices vulnerable to attacks. A tool once termed as secure, will later expose the users to reckless security issues[25]. IoT creators rush their developments, compromising security loopholes. Traditional security updates are no longer applied due to the rush to sell their creations. After the sales are made, it is the end of their support journey. Customers are always advised to act pre- cautionary when choosing these devices. Even amidst exiting features and offers, security should always come first[49].

### 5.4 IoT Malware and Ransomware

IoT malware and ransomware are not a new subject [2]. People are fond of these security issues from the begging of the internet. Malware and ransomware are hybridizing merging to form very complicated attacks [8]. Ransomware attacks focus on disabling or limiting the device's functionalities. The result also works to steal user data from these devices. Ransomware attacks are of different forms. Most of these solicit financial resources for them to stop further persecutions. The structure of accosting threatens the security of data shared across IoT devices. On the other part, malware acts to be malicious programs replicating into the IoT devices[48]. The various forms of malware create significant issues for the IoT platform. Malware could work in multiple types across multiple levels of the IoT interconnections.

### 5.5 Data Security and Privacy

IoT devices involve a large number of data sharing and receiving. This is the most massive issue in IoT security and privacy. The use of IoT requires transactions with extensive data across the internet[23]. The wide array of IoT devices creates the chances of being interfered with during their operations. Due to the number of users across the internet, a great deal of these is irresponsible. They could hijack the flow information across the internet, putting data into a vulnerable state. Data violation is a commonality amidst the broad applications of IoT. Significant companies are tasked with handling big data used in the IoT operations. Due to the demand for user data, they might share it with their counterparts without the user's consent. There is a need to impose compliance and set applicable privacy rules. These efforts are to ensure the anonymity of the data. Cached data and the one no longer needed should undergo secure disposal[24].

## 6 Current Measures in IoT Security and Privacy

In the wake of enormous security loopholes, IoT experts are restless. Therefore, there is a comprehensive review of these devices, attempting to solve their security and privacy issues. While it is clear it is impossible to secure a one hundred percent success rate on security issues, the stakeholders are just working to eliminate the prevalence of these attacks. Various security countermeasures are currently implicated[27][28].

### 6.1 Security Analytics

Evidence suggests the need for performing analytics before embarking on a security option. Security analytics paves the way for analysis and understanding of these patterns. Data is collected, correlated, and analyzed from various centers assisting in the potential threats [3]. Multi-dimensional security analytics continues to pave the way for such developments. Anomalies and malicious patterns are then correlated from several domains. Security agents can correct them, preventing negative implications on the IoT connections.

### 6.2 Protection of Communications

Securing communication is a gateway to enhanced security countermeasures. IoT relies on the interconnection of devices that are highly compromised[48]. Many users are unaware of the risk they are exposed to through the use of IoT devices. Encryption is a way to ensure secure communication between clients. It is a principle ensuring that only authentic accessories get access to the information. Recommended encryptions for these communications include AES 256 and HTTP[9]. Such suites protection preventing impregnable interfaces.

### 6.3 Securing Network

The pathway to all IoT connections is the use of various networks[22]. They use back-end connections to the internet to access and send data. IoT networks are instrumental in fully supporting communications[43]. Securing IoT networks supports smooth operations. Endpoint security features should be incorporated. They include anti-malware, firewalls, and intrusion detection prevention.

### 6.4 Authentication

The use of authentication is beneficial in acknowledging that only authorized people should access **the** information. Vulnerabilities are common in IoT devices without appropriate authentication techniques[4]. Several authentication features are in place today. They include the use of biometrics, digital certificates, and two-factor authentication. These recommendations require personal profiles like fingerprints to be granted access to the systems[29].

| Reference Author and Year | Tittle | Scope and Population | Findings | Critique |
|---|---|---|---|---|
| Abomhara, M., & Køien, G. (2015) | Cyber Security and the Internet of Things: Vulnerabilities, Threats, Intruders, and Attacks | The paper uses a qualitative analysis in the classification of threat types. Analysis and characterization of attacks and IoT intrude | IoT security challenges include confidentiality, privacy, and trust. Security threats are measured in terms of actors, capability, and motivation. | A lot of focus should be emphasized on both end-users and vendors. New IoT standards should address the likely consequences of IoT breaches. Well-advanced techniques in authentication, access control, identity, and trust management should be discussed. |
| Borgohain, T., Kumar, U., & Sanyal, S. (2018) | Survey of Security and Privacy Issues of the Internet of Things | A general survey of IoT security issues is made. Privacy concerns of end-users | Improvised security measures should be adopted before developing new implementations. | The issues highlighted in the paper do not have any solution to the survey of IoT security issues. |
| Iqbal, M., Olaleye, O., & Bayoumi, M. (2016) | A Review on the Internet of Things (IoT): Security and Privacy Requirements and the Solution Approaches | Internet of Things. Security Privacy Issues Wireless Sensor Networks. RFID | Low resources and system architecture remain a challenge in security measures. | End-to-end security layers should be the central theme in security analytics. Secure authentication and encryption |
| | | Authentication Key management | Strengthened security architectures should be incorporated. | techniques should work within these security layers. |
| Jajieh, M. (2016) | Security and Privacy Issues in Internet of Things (IoT) | IoT Security issues | IoT represents a broadband connection aided by Human Machine Interface. | IoT networks act as the doorway to attacks and exploitations. The range of security concerns bears their countermeasures. |
| Joshitta, R., & Arockiam, L. (2018) | Security in IoT Environment: A Survey | Security issues IoT environment Confidentiality Integrity Availability | There is an overall rise of devices connected to the internet. The concept prompts multiple users to connect to numerous information sources using the internet. | Communicating and storing information with multiple devices is a severe security problem. Reaching to secure connections questions the reliability of the IoT market. |
| Maple, C. (2017) | Security and privacy in the internet of things | Internet of things Security Privacy Trust | The high number of devices contributes to security challenges. The evolution of IoT poses additional risks for new creations[16]. | As new developments are in place, we are struggling to keep up with security policies. Security and privacy considerations remain quite unfortunate in these new creations. |
| Moussa, W., & Dimitri, K. (2018) | IoT And Security-Privacy Concerns: A | Internet of Things Privacy Security Mapping study | RFID and other connectivity protocols are massively | End-to-end communications between the devices remain the utmost |
| | Systematic Mapping Study | | connecting objects. | challenge. Privacy, confidentiality, and integrity top the list of these security issues. |
| Perwej, Y., Omer, M., Sheta, O., & Harb. (2019) | The Future of the Internet of Things (IoT) and Its Empowering Technology | IoT SigFox MEMS ZigBee Web of things (WoT) Long-Range (LoRa) MQTT Wireless HART Z-Wave | IoT encompasses intelligent networks concatenating various elements across the internet for communication and exchange of information. | IoT physical objects incorporate various technologies, including actuators, electronics, and sensors. IoT enablers determine the success of the particular device in terms of the workability. |
| Razzaq, M., Qureshi, M., Gill, S., & Ullah, S. (2017) | Security Issues in the Internet of Things (IoT): A Comprehensive Study | Internet of things Security issues in IoT Security Privacy | Wireless communication networks are highly prone to security threats. Deployment of IoT security is a major priority | New improvised IoT devices will rise in the future. Rising devices without security countermeasures provide will raised the susceptibility of security issues. |
| Rouhifar, M., Bahramzadeh, Hedayat, A., & Aghazarian. (2018) | Statistical Analysis on IoT Research Trends: A Survey | Internet of things Trends Statistical analysis Classification Research domains & subdomains | Technology and software rank as the first concentration during IoT creations. Communication and trust management comes as second and third options. | Failure to prioritize communication and trust is a significant security threat. Cloud and WSN technologies are not spared in the developments associated with IoT security impairments. |
| Sedrati, A., & | A Survey of | Internet of things | A combination of | Security issues lie |
| Mezrioui, A. (2018) | Security Challenges in the Internet of Things | Challenges Security Privacy Lightweight | real-life objects and virtual life is becoming a reality. The severity of functions and operations of IoT are at a great risk imposed by security challenges. | within the layered architecture in the available IoT devices. |
| Suo, H., Wan, J., Zou, C., & Liu, J. (2012) | Security in the Internet of Things: A Review | Internet of things Security Privacy Confidentiality Challenges | IoT devices face new challenges as they are still developed. | Critical focuses on IoT should be within communication security, encryption, cryptography, and protection of sensor data. |
| Arjumin, N., Sidek, S., Hassan, M., Kudus, N., Mohamed, S., & Rajikon, M. (2015) | The Challenges and Contribution of the Internet of Things (IoT) for Smart Living | Internet of things Smart living Challenges Contribution | The use of IoT brings better borderless connections in multiple industries. Smart living is a new wave facilitating endless interconnections in the comfort of home. | Smart cities are complicated due to insecure connections. Focuses on smart living should be implemented with a high degree of intelligence and privacy. |
| Arockiam, L., & Mary, P. (2018) | Issues and Challenges hampering the evolution of IoT big data analytics | Big data IoT Sensors Analytics | Big data analytics in IoT interactions is widening. | The use of big data analytics should concentrate on finding the limitations within the IoT devices. |
| Fernandes, P., Monteiro, A., & Lasrado, S. (2016) | Evolution of Internet of Things (IoT): Security Challenges and Future Scope | IoT Breakthrough Computational Necessity Network layer | IoT is purposed for connections all over the world. Lack of human involvement takes | IoT is quite early where failure to encompass individuals creates more challenges for these devices. |

| | | | the exciting bit of these applications. | |
| --- | --- | --- | --- | --- |
| Hussein, A. (2019). Internet of Things (IoT) | Research Challenges and Future Applications | Internet of Things IoT applications IoT challenges Future Technologies Smart cities Smart environment Smart agriculture Smart living | Applications of IoT fit in various domains. More advances in IoT will gradually realize new virtualized realities. | More 'smart' life is bringing in new security complications. People enjoy the enticing status of these elements without understanding the potential risks associated with the creations. |
| Madushanki, A., Halgamuge, M., Wirasagoda, H., & Syed, A. (2019). | Adoption of the Internet of Things (IoT) in Agriculture and Smart Farming towards Urban Greening: A Review | Internet of Things IoT Agricultural Smart farming Business Sensor data Automation | IoT is one of the creations working to maximized agricultural practices. Sensing technologies are widely improvised in agriculture to encourage smart farming. | Sensors, just like any other IoT, gets channeled to an array of security threats. |
| Mankar, C., & Mankar, H. (2016) | Internet of Things (IoT) an Evolution | Ubiquitous Computing | Crucial technologies are in the birth of IoT in the world. Multiple applications are considering the right technology to be induced in their operations. | The choice of IoT technology is full of controversies. Security breaches happen at all levels of technologies, no matter the superiority. Proper adherence to the rules and security guidelines is a major justification. |
| Maple, C. (2017) | Security and | Internet of things | The evolution of | The growth of IoT |

| | privacy in the internet of things | Security Privacy Trust | IoT keeps on shifting rapidly bearing new creations now and then. | is not of all sorts of wellness due to major security breaches in place. |
|---|---|---|---|---|
| Vikas. (2015) | Internet of Things (IoT): A Survey on Privacy Issues and Security | IoT Security Privacy IoT Architecture Encryption Communication | An intelligent collaboration of IoT devices brings a wholesome new level of security issues. | The adaptability of IoT collaborations creates several security breaches. Security and privacy are heavily dependent on user behavior. Do people know what these devices are for? |

## 7  What is the Future of IoT Security and Privacy?

Indeed, the issues facing IoT security and privacy are enormous[18]. It is not quite sure about how the future will look. There is the susceptibility of raised cases of IoT security and privacy issues in the future ac- cording to many security analyses. The pace of these issues is engaging without accuracy in making[10]. More complications in IoT devices will emerge in the future. But on the counter side, security agents will continue to harden their security inputs as possible. Companies will collaborate to strengthen the security of these devices[20]. IoT education continues to emerge as a trend with its potential applications in the future.

## 8  Conclusion

Recent radical transformations of ubiquitous internet of things (IoT) have filled all over the world. These transformations properly oversee multiple capacities where 'things' interact with each other across the web. Such interactions allow data to be generated and analyzed extracting valuable information. Users are experienced to release and receive data at their consent. The high interconnections are the world's Internet of Things (IoT)[46]. Profuse advantages are linked with the increasing usability of IoT devices. Business logicalities, improved people's lives, among other efficiencies are highlighted to mention a few prominently. The list goes on building with new creations happening now and then. However, the altered realities by IoT contain umpteen security and privacy loopholes. Traditional shielding measures are unfit to the new paradigms in technology. Recent standards of IoT are complicated, requiring in-depth security and privacy measures.

# References

1. Abomhara, M., & Køien, G. (2015). Cyber Security and the Internet of Things: Vulnerabilities, Threats, Intruders, and Attacks. Journal of Cyber Security and Mobility, 4(1), 65-88 .
2. Borgohain, T., Kumar, U., & Sanyal, S. (2018). Survey of Security and Privacy Issues of Internet of Things. Retrieved from https://arxiv.org/ftp/arxiv/papers/1501/1501.02211.pdf
3. Iqbal, M., Olaleye, O., & Bayoumi, M. (2016). A Review on Internet of Things (Iot): Security and Privacy Requirements and the Solution Approaches. Global Journal of Computer Science and Technology: E Network, Web Security, 16(7).
4. Jajieh, M. (2016). Security and Privacy Issues in Internet of Things(IoT).
5. Joshitta, R., & Arockiam, L. (2018). Security in IoT Environment: A Survey. Int. Journal of Information Technology Mechanical Engineering, 2(7), 1-8.
6. Maple, C. (2017). Security and privacy in the internet of things. Journal of Cyber Policy, 2(2).
7. Moussa, W., & Dimitri, K. (2018). IOT AND SECURITY-PRIVACY CONCERNS: A SYSTEMATIC MAPPING STUDY. International Journal of Network Security Its Applications, 10(6).
8. Perwej, Y., Omer, M., Sheta, O., & Harb. (2019). The Future of Internet of Things (IoT) and Its Empowering Technology. International Journal of Engineering Science and Computing.
9. Razzaq, M., Qureshi, M., Gill, S., & Ullah, S. (2017). Security Issues in the Internet of Things (IoT): A Comprehensive Study. International Journal of Advanced Computer Science and Applications, 8(6).
10. Rouhifar, M., Bahramzadeh, Hedayat, A., & Aghazarian. (2018). Statistical Analysis on IoT Research Trends: A Survey. J. ADV COMP ENG TECHNOL, 4(2).
11. Sedrati, A., & Mezrioui, A. (2018). A Survey of Security Challenges in Internet of Things. Advances in Science, Technology and Engineering Systems Journa, 3(1), 274-280.
12. Suo, H., Wan, J., Zou, C., & Liu, J. (2012). Security in the Internet of Things: A Review. 2012 International Conference on Computer Science and Electronics Engineering.
13. Arjumin, N., Sidek, S., Hassan, M., Kudus, N., Mohamed, S., & Rajikon, M. (2015). The Challenges and Contribution of Internet of Things (Iot) for Smart Living. International Journal of Recent Technology and Engineering (IJRTE), 8(1).
14. Arockiam, L., & Mary, P. (2018). Issues and Challenges hampering the evolution of IoT big data analytics. International Journal of Applied Engineering Research, 10(82).
15. Fernandes, P., Monteiro, A., & Lasrado, S. (2016). EVOLUTION OF INTERNET OF THINGS (IOT): SECURITY CHALLENGES AND FUTURE SCOPE. International Journal of Latest Trends in Engineering and Technology, 164-168.
16. Hussein, A. (2019). Internet of Things (IOT): Research Challenges and Future Applications. International Journal of Advanced Computer Science and Applications, 10(6).
17. Madushanki, A., Halgamuge, M., Wirasagoda, H., & Syed, A. (2019). Adoption of the Internet of Things (IoT) in Agriculture and Smart Farming towards Urban Greening: A Review. International Journal of Advanced Computer Science and Applications, 10(4), 11-28.

18. Mankar, C., & Mankar, H. (2016). Internet of Things (IoT) an Evolution. International Journal of Computer Science and Mobile Computing, 5(3), 772-775.
19. Maple, C. (2017). Security and privacy in the internet of things. Journal of Cyber Policy, 2(2), 155-184.
20. Vikas. (2015). Internet of Things (IoT): A Survey on Privacy Issues and Security. International Journal of Scientific Research, 1(3), 168.
21. Ch, R., Gadekallu, T. R., Abidi, M. H., Al-Ahmari, A. (2020). Computational System to Classify Cyber Crime Offenses Using Machine Learning. Sustainability, 12(10), 4087.
22. Maddikunta, P. K. R., Srivastava, G., Gadekallu, T. R., Deepa, N., Boopathy, P. (2020). Predictive model for battery life in IoT networks. IET Intelligent Transport Systems.
23. RM, S. P., Maddikunta, P. K. R., Parimala, M., Koppu, S., Reddy, T., Chowdhary, C. L., Alazab, M. (2020). An effective feature engineering for DNN using hybrid PCA-GWO for intrusion detection in IoMT architecture. Computer Communications.
24. Deepa, N., Prabadevi, B., Maddikunta, P. K., Gadekallu, T. R., Baker, T., Khan, M. A., Tariq, U. (2020). An AI based intelligent system for healthcare analysis using Ridge Adaline Stochastic Gradient Descent Classifier. Journal of Supercomputing.
25. Maddikunta, P. K. R., Gadekallu, T. R., Kaluri, R., Srivastava, G., Parizi, R. M., Khan, M. S. (2020). Green communication in IoT networks using a hybrid optimization algorithm. Computer Communications.
26. RM, S. P., Bhattacharya, S., Maddikunta, P. K. R., Somayaji, S. R. K., Lakshmanna, K., Kaluri, R., ... Gadekallu, T. R. (2020). Load balancing of energy cloud using wind driven and firefly algorithms in internet of everything. Journal of Parallel and Distributed Computing.
27. Iwendi, C., Jalil, Z., Javed, A. R., Reddy, T., Kaluri, R., Srivastava, G., Jo, O. (2020). KeySplitWatermark: Zero Watermarking Algorithm for Software Protection Against Cyber-Attacks. IEEE Access, 8, 72650-72660.
28. Numan, M., Subhan, F., Khan, W. Z., Hakak, S., Haider, S., Reddy, G. T., ... Alazab, M. (2020). A systematic review on clone node detection in static wireless sensor networks. IEEE Access, 8, 65450-65461.
29. Patel, H., Singh Rajput, D., Thippa Reddy, G., Iwendi, C., Kashif Bashir, A., Jo, O. (2020). A review on classification of imbalanced data for wireless sensor networks. International Journal of Distributed Sensor Networks, 16(4), 1550147720916404.
30. Reddy, T., RM, S. P., Parimala, M., Chowdhary, C. L., Hakak, S., Khan, W. Z. (2020). A deep neural networks based model for uninterrupted marine environment monitoring. Computer Communications.
31. Iwendi, C., Maddikunta, P. K. R., Gadekallu, T. R., Lakshmanna, K., Bashir, A. K., Piran, M. J. (2020). A metaheuristic optimization approach for energy efficiency in the IoT networks. Software: Practice and Experience.
32. Bhattacharya, S., Kaluri, R., Singh, S., Alazab, M., Tariq, U. (2020). A Novel PCA-Firefly based XGBoost classification model for Intrusion Detection in Net- works using GPU. Electronics, 9(2), 219.
33. Azab, A., Layton, R., Alazab, M., Oliver, J. (2014, November). Mining malware to detect variants. In 2014 Fifth Cybercrime and Trustworthy Computing Conference (pp. 44-53). IEEE.
34. Alazab, M., Layton, R., Broadhurst, R., Bouhours, B. (2013, November). Malicious spam emails developments and authorship attribution. In 2013 Fourth Cybercrime and Trustworthy Computing Workshop (pp. 58-68). IEEE.

35. Alazab, M., Venkatraman, S., Watters, P., Alazab, M. (2013). Information security governance: the art of detecting hidden malware. In IT security governance innovations: theory and research (pp. 293-315). IGI Global.
36. Farivar, F., Haghighi, M. S., Jolfaei, A., Alazab, M. (2019). Artificial Intelligence for Detection, Estimation, and Compensation of Malicious Attacks in Nonlinear Cyber-Physical Systems and Industrial IoT. IEEE transactions on industrial informatics, 16(4), 2716-2725.
37. Azab, A., Alazab, M., Aiash, M. (2016, August). Machine learning based botnet identification traffic. In 2016 IEEE Trustcom/BigDataSE/ISPA (pp. 1788-1794). IEEE.
38. Alazab, M., Huda, S., Abawajy, J., Islam, R., Yearwood, J., Venkatraman, S., Broadhurst, R. (2014). A hybrid wrapper-filter approach for malware detection. Journal of networks, 9(11), 2878-2891.
39. Garg, S., Singh, A., Batra, S., Kumar, N., Yang, L. T. (2018). UAV-empowered edge computing environment for cyber-threat detection in smart vehicles. IEEE Network, 32(3), 42-51.
40. Bali, R. S., Kumar, N. (2016). Secure clustering for efficient data dissemination in vehicular cyber–physical systems. Future Generation Computer Systems, 56, 476-492.
41. Chaudhary, R., Kumar, N., Zeadally, S. (2017). Network service chaining in fog and cloud computing for the 5G environment: Data management and security challenges. IEEE Communications Magazine, 55(11), 114-122.
42. Singh, A., Maheshwari, M., Kumar, N. (2011, April). Security and trust management in MANET. In International Conference on Advances in Information Technology and Mobile Communication (pp. 384-387). Springer, Berlin, Heidelberg.
43. Vora, J., Italiya, P., Tanwar, S., Tyagi, S., Kumar, N., Obaidat, M. S., Hsiao, K. F. (2018, July). Ensuring privacy and security in E-health records. In 2018 International conference on computer, information and telecommunication systems (CITS) (pp. 1-5). IEEE.
44. Hathaliya, J. J., Tanwar, S., Tyagi, S., Kumar, N. (2019). Securing electronics healthcare records in healthcare 4.0: a biometric-based approach. Computers Electrical Engineering, 76, 398-410.
45. Krishnasamy, L., Dhanaraj, R. K., Ganesh Gopal, D., Reddy Gadekallu, T., Aboudaif, M. K., Abouel Nasr, E. (2020). A Heuristic Angular Clustering Framework for Secured Statistical Data Aggregation in Sensor Networks. Sensors, 20(17), 4937.
46. Mohan Vijay (2018). AN UPDATED NEW SECURITY ARCHITECTURE FOR IOT NETWORK BASED ON SOFTWARE-DEFINED NETWORKING (SDN). IRJCS:: International Research Journal of Computer Science, Volume V, 77-81.
47. Krishna Kagita, M. (2019). Security and Privacy Issues for Business Intelligence in a IoT. In Proceedings of 12th International Conference on Global Security, Safety and Sustainability, ICGS3 2019 [8688023]
48. Krishna Kagita, M. and M. Varalakshmi, 2020. A detailed study of security and privacy of Internet of Things (IoT). International Journal of Computer Science and Network, 9 (3): 109–113.
49. Navod Neranjan Thilakarathne, Mohan Krishna Kagita, Dr. Thippa Reddy Gadekallu. (2020). The Role of the Internet of Things in Health Care: A Systematic and Comprehensive Study. International Journal of Engineering and Management Research, 10(4), 145-159.